\documentclass[prd,amsfonts,twocolumn,nofootinbib,showpacs]{revtex4-1}
\usepackage{graphicx}
\usepackage{dcolumn}
\usepackage{bm}
\usepackage{amsmath}
\usepackage{amsfonts} 
\usepackage{amssymb}
\usepackage{aas_macros}

\def\fgas{{$f_{gas}$\ }}
\def\fgasm{{\rm f_{gas }}}

 \def\fom{{FoM }}

\begin{document}

\title{Optimizing Observational Strategy for Future $f_{\rm gas}$ Constraints}

\author{Silvia Galli$^{a,b,c,d}$, James G. Bartlett$^{a,e}$, Alessandro Melchiorri$^b$}
\affiliation{$^a$ Astroparticule et Cosmologie, CNRS (UMR 7164), Universit\'e Denis Diderot Paris 7, B{\^a}timent Condorcet, 
	10 rue A. Domon et L. Duquet, Paris, France}
\affiliation{$^b$ Physics Department and INFN, Universita' di Roma ``La Sapienza'', Ple Aldo Moro 2, 00185, Rome, Italy}
\affiliation{$^c$ UPMC Univ Paris 06, UMR7095, Institut d'Astrophysique de Paris, F-75014, Paris, France}
\affiliation{$^d$ CNRS, UMR7095, Institut d'Astrophysique de Paris, F-75014, Paris, France}
\affiliation{$^e$ Jet Propulsion Laboratory, California Institute of Technology, 4800 Oak Grove Drive, Pasadena, California, USA }
\pacs {26.35.+c, 98.80.Cq, 98.80.Ft \hfill SACLAY--T09/046}

\date{March 23, 2011}

\begin{abstract}
The Planck cluster catalog is expected to contain of order a thousand galaxy clusters, both newly discovered and previously known, detected through the Sunyaev-Zeldovich effect over the redshift range $0 \lesssim z \lesssim 1$.   Follow-up X-ray observations of a dynamically relaxed sub-sample of newly discovered Planck clusters will improve constraints on the dark energy equation-of-state found through measurement of the cluster gas mass fraction ($f_{gas}$).  In view of follow-up campaigns with XMM-Newton and Chandra, we determine the optimal redshift distribution of a cluster sample to most tightly constrain the  dark energy equation of state.   The distribution is non-trivial even for the standard $w_{\rm 0}$-$w_{\rm a}$ parameterization. We then determine how much the combination of expected data from the Planck satellite and \fgas data will be able to constrain the dark energy equation-of-state. Our analysis employs a Markov Chain Monte Carlo method as well as a Fisher Matrix analysis. We find that these upcoming data will be able to improve the figure-of-merit by at least a factor two.
\end{abstract}

\maketitle
\section*{Introduction}
\label{intro}

Thanks to the detailed modeling of galaxy clusters afforded by X-ray observations from 
XMM-Newton and Chandra, measurements of the cluster gas mass fraction (\fgas) have become a 
viable cosmological probe \cite{allen,rapetti,ettori,laroque} (see also \cite{white93,pen97,sasaki96}).  
Effectively a distance measurement,  \fgas determinations as a function of redshift are a geometric probe of dark energy and,
when combined with cosmic microwave background (CMB) anisotropy  data from the WMAP satellite \cite{wmap7}, they 
have produced valuable constraints on key dark energy parameters.  Using $42$ clusters, for example, \cite{allen} place a constraint on the 
dark energy equation-of-state parameter $P=w_0 \rho$  of $w_0=-1\pm0.14$ at the $68\%$ confidence level.
One can substantially improve this constraint by applying the technique to a larger  number of clusters 
(e.g., $\sim 400$) and shed new light on dark energy \cite{rapetti}.  

To reduce the impact of modeling uncertainties, one preferentially uses massive and dynamically relaxed clusters,
because gravity dominates their gas physics and deviations from hydrostatic equilibrium should be small.  
This, however, limits the number of available systems.  Increasing the sample size of suitable systems 
requires the surveying of large volumes.  The vast majority of samples used to date come from the ROSAT All-Sky 
survey (RASS) and extensive compilations from the mission archive.  

Surveys based on the Sunyaev-Zeldovich (SZ) effect (\cite{sunyaev72}, see 
also the reviews by \cite{birkinshaw99,carlstrom02}) deliver a powerful capability 
of building new cluster samples.  
Cluster SZ surveys now produce catalogs of several tens of clusters \cite{act,lueker10} to over one hundred \cite{Planck2011-5.1a},
 and sample size will dramatically increase in the near future as more data is analyzed and sky covered.  
Of particular interest in the present context, the Planck satellite \cite{planckmission} is constructing the first all-sky 
cluster catalog since ROSAT. Thanks to its large survey volume, Planck will increase the 
number of hot, X-ray luminous and massive systems available for study at intermediate redshifts, beyond the RASS limit, and out to $z\sim 1$. It is expected to detect of order a thousand galaxy clusters, both newly discovered and previously known \cite{chamballu10}.   The Early SZ (ESZ) cluster list of the highest significance 
Planck detections over the complete sky was recently published \cite{Planck2011-5.1a}
along with a set of Planck SZ results \cite{Planck2011-5.1b,Planck2011-5.2a,Planck2011-5.2c}.

In view of planned X-ray follow-up campaigns, we examine the potential
gains in dark energy constraints from \fgas measurements over 
these new cluster samples.  In particular, we quantify the optimal redshift 
distribution for target samples.  To this end, we employ the standard Figure of Merit (\fom) 
defined by the Dark Energy Task Force (DETF) \cite{detf} for a redshift
dependent equation of state parametrized as $w(a)=w_{\rm 0}+w_{\rm a}(1-a)$\cite{chevallier00,linderw0w1,linderhuterer}.  

The \fom is proportional to the inverse of the area contained in the
95\% confidence contour of the $w_{\rm 0}-w_{\rm a}$ plane. 
The DETF showed that this area is equal to the area enclosed in the $w_{\rm a}-w_{\rm p}$ plane, where $w_{\rm p}$ is the value of $w(a)$ at the pivot scale factor $a_p$ where its uncertainty $\sigma(w,a)$ is
minimized \cite{linderpivot}. It can be easily shown that the pivot scale factor is:
\begin{eqnarray}
 a_p&=&\frac{cov(w_0,w_a)}{\sigma^2(w_a)}+1
\end{eqnarray}
and thus the error on $w_p=w(a_p)$ is:
\begin{eqnarray}
\sigma^2(w_p)&=&\sigma^2(w_0)-\frac{cov(w_0,w_a)}{\sigma^2(w_a)}
\end{eqnarray}
where $cov(w_0,w_a)$ is the covariance between $w_0$ and $w_a$.
As there is little correlation, at least in the Gaussian approximation, between $w_{\rm a}-w_{\rm p}$, the area is simply proportional to the product of the standard deviations: $\sigma(w_{\rm p}) \times \sigma(w_{\rm a})$.

\noindent This motivates the definition:

\begin{equation}
{\rm FoM}=[\sigma(w_{\rm p})\times\sigma(w_{\rm a})]^{\rm -1}.
\label{eq:fom}
\end{equation}

In this paper we look for the optimal cluster redshift distribution that maximizes the \fom for a combined
future CMB+\fgas data set.  We consider as illustration an observational X-ray campaign with a total available observational time of $\sim 5 Ms$ to measure \fgas of newly discovered clusters or to improve the \fgas determinations of previously observed clusters.  We assume that with this amount of time it will be possible to obtain \fgas measurements for $\sim 100$ clusters with an error of $\sim$ 10\% and with a maximum redshift of $z_{max}=1$.  This scenario is motivated in particular by the expectation that the Planck clusters will be luminous X-ray objects \cite{chamballu10}, as it is also suggested by the preliminary follow-up of the ESZ \cite{Planck2011-5.1b}.  Such an increase in sample size will certainly auger an improvement on dark energy constraints.
Our study includes the CMB temperature and polarization data expected from Planck \cite{planck}, together with the cluster \fgas measurements.
Using the Fisher matrix formalism, we forecast uncertainties on dark energy parameters 
and calculate the \fom  given different cluster redshift distributions.  We determine the optimal distribution as that which maximizes the \fom.

Once the optimal redshift distribution has been identified, we perform a Monte Carlo analysis on simulated data 
and examine the improvement in constraining dark energy parameters when measuring \fgas.   

Our paper is organized as follows: in section \ref{sec:fgas} we briefly describe the \fgas method and the dependence of the data on cosmological parameters. Section \ref{sec:fisher} describes the Fisher Matrix method and its use for \fgas and CMB data, while section \ref{sec:optimal} presents the analysis of the optimal cluster redshift distribution. Finally, in section \ref{section:mcmc} we present parameter constraints from the Monte-Carlo study. 
In section \ref{sec:conclusions}, we discuss our conclusions.

\section{The $f_{gas}$ method}
\label {sec:fgas}

We begin by briefly recalling the \fgas method and its use for 
cosmological analyses. For greater detail, we refer the reader to \cite{white93,evrard}.

The X-ray gas mass fraction \fgas is defined as the ratio of
the X-ray emitting gas mass to the total mass of a cluster. It is 
determined from the observation of the X-ray surface brightness
and the spectrally-determined, deprojected gas temperature profile.  
As both spherical symmetry and hydrostatic equilibrium are assumed for the clusters considered  \cite{allen}, the analysis is best limited to the hottest, dynamically relaxed clusters. 

Since clusters form from matter initially distributed over large volumes, we expect their composition to reflect that of the cosmic mean \cite{white93}.  The cluster baryon fraction $f_b$, defined as the total baryonic mass over the total cluster mass, should faithfully reflect the cosmic 
baryon fraction $$f_b=f_{gas}^{\rm true}+f_{stars}\propto\Omega_b/\Omega_m,$$ 
where $f_{stars}$ is the fraction of baryons in stars and $f_{gas}^{\rm true}$ is the true value of the 
cluster gas mass fraction.  This relation is supported by hydrodynamical simulations \cite{Eke,Nagai,Crain}.   Turning the relation around, allowing for systematic and astrophysical effects \cite{allen,rapetti} and expressing the cosmic baryon fraction through the physical densities $\omega_{b,m}=\Omega_{b,m}h^2$, with $h$ being the reduced Hubble parameter $h=H_0/(100 {\rm \,Km/s/Mpc})$,  we write:
\begin{equation}
\label{systematics2}
  f^{\rm true}_{\rm gas}(z)=\gamma\,K\,\left(\frac{\omega_{\rm b}}{\omega_{\rm m}}\right)\left(\frac{b(z)}{1+s(z)}\right).
\end{equation}
where $\gamma$ quantifies potential departures from hydrostatic equilibrium, e.g., due to non-thermal pressure support;
$K$ is a normalization uncertainty related to instrumental calibration and
certain modelling issues; $b(z)=b_{\rm 0}(1+\alpha_{\rm b}z)$ represents a possible cluster depletion factor and its evolution, and
$s(z)=s_{\rm 0}(1+\alpha_{\rm s}z)$ is the stellar baryon fraction as a function of redshift (see, e.g., \cite{allen}).

When calculating \fgas from the X-ray observations, one needs to adopt a reference cosmological model, which is usually taken to be $\rm \Lambda CDM$, in order, e.g., to relate the observed flux to the luminosity of the  bremsstrahlung emission, related to the density of the emitting gas. It can be shown that the mass of the gas estimated from the deprojected X-ray surface brightness has a $M_{gas}\propto d_A^{5/2}$ dependence on the angular diameter distance, while the estimated total mass has a dependence of the type $M\propto d_A$ \cite{sasaki96}. Thus, the \fgas has a $\propto d_A^{3/2}$ dependence on the angular diameter distance.
 If the reference cosmology is not the true one, the gas fraction found by adopting the reference cosmology, $f^{\rm ref}_{\rm gas}$ (the {\em observed fraction}), is related to the true gas fraction via:

\begin{equation}
  f^{\rm ref}_{\rm gas}(z;\theta^{\rm ref }_{2500})=f^{\rm true}_{\rm gas}(z;\theta^{\rm true}_{2500})\left[\frac{d^{\rm ref }_{A}(z)}{d^{\rm true}_{A}(z)}\right]^{1.5}\left[\frac{\theta^{\rm ref }_{2500}(z)}{\theta^{\rm true}_{2500}(z)}\right]^{\eta}\,.
\label{eq:fgas}
\end{equation}
Here, we suppose the gas mass fraction is measured within an angular radius $\theta_{2500}$ corresponding to a physical radius $r_{2500}$\footnote{The radius $r_{2500}$ is defined as that radius within which the mean density is 2500 times the critical density of the universe at the redshift of the cluster.}. If the true cosmology is different from the reference one, the misestimate of $\theta_{2500}$ introduces an additional dependence on cosmology, due to the fact that $f^{\rm true}_{\rm gas}$ smoothly varies with radius \cite{Eke}, $f_{gas}\propto (r/r_{2500})^\eta$. It can be shown that $\theta_{2500}$ depends on the Hubble parameter and on the angular diameter distance as $\theta_{2500}=r_{2500}/d_A(z)\propto (H(z)d_A(z))^{-1}$ \cite{rapetti}. The parameter $\eta$ is thus taken into account along with potential systematic effects.  The angular diameter distance is defined as $d_A(z)= (1+z)^{-1} S_k(\chi)$. $S_k$ depends on cosmological parameters as:
 \begin{equation}
S_k(\chi)=\left\{
\begin{array}{ll}
\frac{1}{\sqrt{|\Omega_k|}} \sin{\chi \sqrt{|\Omega_k|}} & k > 0\\
\chi & k = 0\\
\frac{1}{\sqrt{|\Omega_k|}}\sinh{\chi\sqrt{|\Omega_k|}} & k <0
\end{array}\right.
\end{equation}
where $k$ is the curvature of the universe and $\Omega_k h^2=h^2-\omega_m-\omega_{de}$ is the curvature parameter, with $\omega_m$ and $\omega_{de}$ the physical energy
density parameters in dark matter and dark energy, respectively.  The comoving distance $\chi$ is defined as:
\begin{equation}
  \chi(z)=\int_{0}^{z}\frac{dz'}{H(z')}
\end{equation}

\noindent where $H(a)$ is the Hubble parameter that evolves with redshift according to Friedmann's equation:

\begin{equation} 
\left[\frac{H(z)}{100 {\rm Km/s/Mpc}}\right]^2=\omega_m (1+z)^{3}+\Omega_k h^2 (1+z)^{2}+\omega_{de}g(z)
\end{equation}

\noindent where $g(z)$ determines the evolution of the dark energy density.
The influence of dark energy enters herein.  For a dark energy equation-of-state parameterized 
as $w(z)=w_{\rm 0}+w_{\rm a}(1-(1+z)^{-1})$, we have:
\begin{eqnarray}
 g(a)&=&e^{\int_{0}^{z}-3 (w(z)+1)\frac{dz}{1+z}} \nonumber\\
&=&(1+z)^{3(1+w_0+w_a)}e^{-3w_a(1-(1+z)^{-1})}
\label{eq:ga}
\end{eqnarray}

Equations \ref{systematics2} and \ref{eq:fgas} describe the dependence of the observed $f^{\rm ref}_{\rm gas}$  on the actual cosmological parameters, which we constrain by fitting to the observed redshift evolution of $f^{\rm ref}_{\rm gas}$.
{Unfortunately, there are several degeneracies between parameters that cannot be broken using  $f_{gas}$ only data; e.g., systematic effect parameters such as $\gamma$ or $\kappa$, are almost completely degenerate with $\omega_b$. Thus, priors or additional complemetary data are required.}

\begin{figure}[h!]
\begin{center}
\includegraphics[width=0.5\textwidth]{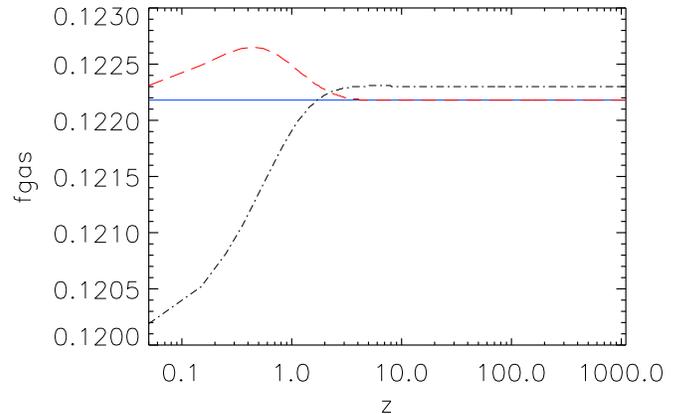}
\caption{Evolution of \fgas with redshift for different values of the true cosmological parameters. The blue solid line shows the evolution for true cosmology parameters equal to the reference values, as reported in Tab. \ref{tab:fiducial} and systematics as in Tab. \ref{tab:systematics}. The dashed red line is for $w_0=-0.975$ and $w_a=-0.09$ in the true cosmology, while the dot-dashed is for $w_0=-0.975$ and $h_0=0.71$. Each of these three sets of parameters provide similar angular diameter distances at high redshifts.}
\label{paramsonfgas}
\end{center}
\end{figure}

{We therefore include in our analysis priors on systematic effects, and combine the simulated $f_{gas}$ data with mock CMB data of the Planck satellite. In fact, Planck is expected to measure the physical energy density of baryons and of dark matter to the subpercent level, as well as the angular diameter distance at redshift $1100$ with very high accuracy, thus providing information on a combination of $H_0$, $w_0$, $w_a$ and $\Omega_k$. 
These parameters are, however, completely degenerate with each other in CMB data. For this reason, \fgas data, as well as other redshift-distance relation probes like Supernovae \cite{hutturn_optSN1,hutturn_optSN2}, are crucial for dark energy studies.} As an example, Fig. \ref{paramsonfgas} shows the evolution of \fgas for different combinations of the true values of $H_0$, $w_0$ and $w_a$. The combinations are chosen to provide the same angular diameter distance at the redshift of recombination, i.e., to be completely degenerate in CMB data. Is is evident from this figure that the \fgas data breaks this degeneracy.

\section{Fisher matrix Analysis}

\label{sec:fisher}

We expect the Planck cluster catalog to significantly increase the number of massive systems known at intermediate redshifts \cite{chamballu10}, and hence the number of suitable sources for \fgas studies.  The context guiding our present study is the possibility of observing $\sim 100$ clusters with XMM-Newton and Chandra from the newly-detected Planck clusters.  We therefore aim to quantify the gain in cosmological constraints achievable with such a data set.  To that end,  we employ the Fisher Matrix methodology (see e.g., \cite{tegmark}). 
We will also study constraints from \fgas measurements  combined with the CMB anisotropy data expected from Planck.

For the cosmological parameters, we adopt the physical baryon and CDM densities, $\omega_b=\Omega_bh^2$ and
$\omega_c=\Omega_ch^2$, the Hubble parameter $H_0$ and the parameters describing the dark energy equation-of-state (EoS).  We consider a constant EoS, $w$, as well as an evolving EoS of the form $w_{\rm 0}$ and $w_{\rm a}$: $w(z)=w_{\rm 0}+(1-(1+z)^{-1})w_{\rm a}$. 
Furhermore, we examine both the cases of a flat universe and universes with free curvature $\Omega_k$.

When simultaneously fitting for \fgas and CMB data, we add in addition the normalization, $A_s(k=0.002/Mpc)$, and scalar spectral index, $n_{s}$, of the primordial fluctuations, as well as the optical depth to reionization, $\tau$.  As fiducial model for this parameter set, we use the best fit from the 5-year release of the WMAP satellite data (WMAP5) \cite{wmap5}, reported in Table \ref{tab:fiducial}. We verified that using the best fit from the 7-year release (WMAP7) \cite{wmap7} does not change any of our results.
Finally, we also consider seven additional parameters to describe systematic effects as in Eq.~\ref{systematics2} and Eq.~\ref{eq:fgas}, and listed in Table \ref{tab:systematics} with the assumed priors. 
The priors adopted are the same as in \cite{allen}. They reflect the current knowledge of their magnitude and therefore are conservative expectations for future data. Since the Fisher analysis only allows Gaussian priors, for those parameters with a uniform prior, we choose to interpret half of the uniform intervals as the $2\sigma$ bounds of Gaussian priors.
In section \ref{section:mcmc}, we perform a MCMC analysis using the correct priors.

\begin{table}
\begin{center}\begin{tabular}{ |l |l |  }
\hline \hline
Parameter &Fiducial\\\hline
$H_{0}$ &$71.9$ {Km/s/Mpc}  \\ 
$\Omega_b$&$0.02273$\\
$\Omega_c$&$0.1099$\\
$\Omega_k$&$0$\\
$w_{\rm 0}$&$-1$\\
$w_{\rm a}$&$0$\\
$\tau$&$0.089$\\
$n_s$&$0.963$\\
$A_s$&$2.41\times10^{-9}$\\ 
\hline
\hline
\end{tabular}
  \caption{Fiducial WMAP5 marginalized cosmological parameters \cite{wmap5}.}
\label{tab:fiducial}
\end{center}
\end{table} 

For the \fgas data, the Fisher Matrix is:
\begin{equation}
 F^\fgasm_{ij}=\sum_{k=1}^{\rm N_{clusters}} \frac{\partial \fgasm(z_k) }{\partial\theta_i} [\mathcal{C}_{kk}]^{-1} \frac{\partial \fgasm(z_k) }{\partial\theta_j}
\label{fisherfgas}
\end{equation}

 where the sum is over the number of clusters $N_{\rm clusters}$ and $\mathcal{C}$ is the covariance matrix of \fgas. 
 In our case, $\mathcal{C}$ is diagonal and equal to $\mathcal{C}= \sigma I$, with $\sigma= 10\%\times \fgasm(z_k)$, as the errors on the measured \fgas at different redshifts are not correlated and we suppose that the errors are the same for all clusters.

Dividing the redshift range into $N_{bins}$, each containing $n_{bin}$ clusters, Equation \ref{fisherfgas} can be expressed as:
\begin{equation}
 F^\fgasm_{ij}=\sum_{{bin}=1}^{N_{bins}} n_{bin} \frac{\partial \fgasm(z_{bin}) }{\partial\theta_i} \frac{1}{\sigma^2} \frac{\partial \fgasm(z_{bin}) }{\partial\theta_j}
\label{fisherfgasbins}
\end{equation}
 This equation shows explicitly the dependence of the \fgas Fisher Matrix on the redshift distribution of the clusters, i.e., the number $n_{bin}$ of clusters in each bin.  This is what we want to optimize. 

It is interesting to note that the tightest contraints on cosmological parameters do not necessarly follow from clusters at redshifts where the variation of \fgas with the parameters, i.e., $\frac{\partial \fgasm(z_{bin}) }{\partial\theta_i}$, is largest.  In fact, degeneracies between parameters invalidate this deduction, making the optimal redshift distribution depend on the precision of the data and its degeneracy-breaking power, as well as on the influence of other (non \fgas) constraints.

\begin{table}
\begin{center}
\begin{tabular}{ l c| l| c }
\hline \hline

Parameter& & &      \\
\hline                                              
Calibration/Modeling            & $K$              & 1.$\pm$ 0.10& Gaussian         \\        
Non-thermal pressure             & $\gamma$         & $0.9<1<1.1$ & Uniform         \\
Gas depletion              & $b_0$            & $0.66<0.83<1$  & Uniform      \\
$b(z)=b_0 (1+\alpha_{\rm b}z)$   & $\alpha_{\rm b}$ & $-0.1<0.<0.1$& Uniform  \\
Stellar mass              & $s_0$            & ($0.162\pm0.049)h_{70}^{0.5}$ &   Gaussian    \\
$s(z)=s_0 (1+\alpha_{\rm s}z)$     & $\alpha_{\rm s}$ & $0.2<0<0.2$& Uniform \\
Angular correction&$\eta$                 & $0.214\pm0.022$ &Gaussian\\
\hline \hline    
\end{tabular}
\caption{Fiducial values and priors for systematic parameters in \fgas. In the Fisher Matrix analysis, these priors are all considered as Gaussian priors. See text for references and details.}
\label{tab:systematics}
\end{center}
\end{table}

For the CMB data, the Fisher Matrix is:
\begin{equation}
 F^{\rm CMB}_{ij}=\sum_{XY}\sum_{\ell} \frac{\partial {C}_\ell^{X} }{\partial\theta_i} [\mathcal{C}_\ell^{XY}]^{-1} \frac{\partial C_\ell^{Y} }{\partial\theta_j}
\label{fishercmb} 
\end{equation}
where $C^{X,Y}_\ell$ denotes the CMB angular power spectra X,Y=TT, EE, TE, plus the effective noise power spectrum of Equation \ref{noisecmb} below, and $\mathcal{C}_\ell^{XY}$ is the covariance matrix of the $C^{X,Y}_\ell$ as defined in \cite{verde}.
We simulate angular power spectra using the \texttt{camb} code\cite{camb} adopting the expected values for the 143 GHz channel of the Planck satellite \cite{planck}. The experimental noise on the angular power spectrum is described as in \cite{knox}:

\begin{equation}
\label{noisecmb}
N_{\ell} = \left(\frac{w^{-1/2}}{\mu{\rm K\mbox{-}rad}}\right)^2 
\exp\left[\frac{\ell(\ell+1)(\theta_{\rm FWHM}/{\rm rad})^2}{8\ln 2}\right],
\end{equation}

with  $w^{-1}=3\times10^{-4}\mu K^2$ as the temperature noise level (we consider a factor $\sqrt{2}$ larger for polarization noise) and  $\theta_{\rm FWHM}=7'$ for the beam size \cite{planck}.  We neglect the contribution of primordial B-mode polarization spectra and the lensing signal and we take $f_{sky}=1$ for the sky coverage. We however verified that assuming a smaller sky coverage, e.g. $f_{sky}=0.65$, lowers the \fom by  less than $\lesssim 10\%$.

\section{Search for the Optimal distribution}
\label{sec:optimal}

We calculate the total Fisher Matrix to estimate the cosmological parameters by summing the CMB and \fgas\ Fisher Matrices:
$$F^{\rm Tot}(n_{bin})=F^{\rm CMB}+F^\fgasm(n_{bin})$$ 
The number of clusters in each redshift bin $n_{bin}$ are the free parameters in our analysis.
We search for the redshift distribution $n_{bin}$ that provides the highest value for the \fom, calculating the latter from the standard deviations $\sigma_{ii}=(F^{Tot
})_{ii}^{-1}$ for each tested redshift distribution and using a maximization algorithm \cite{praxis}.

We first optimize the cluster distribution for the case of linear evolution of the dark energy equation-of-state in a flat universe,
i.e., when $w(z)=w_{\rm 0}+(1-(1+z)^{-1})w_{\rm a}$, constrained using a combination of CMB and \fgas data.
In Table \ref{optdistribution}, we report the optimal distribution for 100 clusters in 10 bins over the redshift range $z=0-1$.  This  maximizes the \fom to a value of $15.2$.

\begin{table}[h!t]
\begin{center}
\begin{tabular}{  c| c}
\hline \hline

Redshift Interval&Number of clusters $n_{bin}$\\
\hline
0.0 - 0.1  & 38 \\ 
0.1 - 0.2  & 0 \\ 
0.2 - 0.3  & 0 \\ 
0.3 - 0.4  & 45 \\ 
0.4 - 0.5  & 0 \\ 
0.5 - 0.6  & 0 \\ 
0.6 - 0.7  & 0 \\ 
0.7 - 0.8  & 0 \\ 
0.8 - 0.9  & 0 \\ 
0.9 - 1.0 & 17 \\
\hline \hline
\end{tabular}
\caption{Optimal distribution for 100 clusters in the redshift range $z=0-1$ for the ($w_0-w_a, \Omega_k=0$) model.}
\label{optdistribution}
\end{center}
 \end{table}

The distribution presents a peak in the first bin at the lowest redshift, a second peak at an intermediate redshift of $z\sim0.35$ and then a third smaller peak in the last bin. This distribution improves the figure of merit by 30\% compared to a simple uniform distribution, which has a \fom of 11.3.

{The shape of this optimized distribution is determined by the various factors. The expected Planck CMB data provide excellent contraints on the single values of $\omega_b$ and $\omega_c$ at subpercent level. However, they can only constrain the combination of $H_0$, $w_0$, $w_a$ and $\Omega_k$ (but we are here first considering the case of a flat universe) because of the geometric
degeneracy. Thus, the \fgas distribution is  optimized to break at best the degeneracy between these parameters, as well as the additional degeneracies introduced by  the presence of \fgas systematics.

Suppose, at first, that systematics are perfectly known. In this case, measuring the \fgas at low redshift permits us to measure $H_0$, as the effect of $w_0$ and $w_a$ at these redshifts is negligible. This can be easily shown from Eq. \ref{eq:fgas}, which in the limit of small $z$ becomes:}

\begin{equation}
  \lim_{z\to0} f^{\rm ref}_{\rm gas}(z;\theta^{\rm ref }_{2500})= f^{\rm true}_{\rm gas}(z) \left[\frac{H_0^{\rm true }}{H_0^{\rm ref}}\right]^{1.5}
\label{eq:fgasz0}
\end{equation}

\noindent Thus, a measurement of \fgas at the smallest possible redshift separates the effect of $H_0$ from $w_0,w_a$, provided that $\omega_b$ and $\omega_c$ are sufficiently well measured by the CMB data. 

{Consider now the effect of redshift-independent systematics. They can change the value of the \fgas at low redshifts, thus being degenerate with $H_0$. Measuring \fgas at high redshifts can help disentangle these effects, given that the redshift-dependent systematic parameters are limited by priors.
 This explains why the optimal distribution places clusters in the highest redshift bin.} 

{Finally, the central peak isolates the effects of $w_0$ and $w_a$. This peak is expected to be near to or at slightly higher redshift than the equivalence between dark energy and  matter in the fiducial true cosmology, i.e., around $z_{eq}\sim0.42$. At this redshift, the variation of \fgas on true values of $w_0$ and $w_a$ differs from the reference ones is maximized, as also shown in Fig. \ref{paramsonfgas}}.
  
{There are two issues with this result. The first is that the optimal distribution in Tab. \ref{optdistribution} peaks in a bin corresponding to redshifts smaller than $z=0.42$. This is due to systematic effects, as their uncertainties grow with redshift, so measurements at low $z$ can better constrain the parameters of interest. The second is the fact that we obviously do not know the true value of $z_{eq}$; however,  recent data (see e.g. \cite{pandolfi}) show that the  redshift of equivalence is between $0.35  \lesssim z\lesssim 0.55$ at 68\% c.l., in the case of a dark energy equation of state as considered here.
We verified that even if the true $z_{eq}$ is different from the fiducial one, within the current constraints, the optimal distribution of Tab. \ref{optdistribution} still provides a FOM that is very close (within $\sim 5\%$) to the maximum value.

Thus we conclude that the use of the optimal distribution of Tab. \ref{optdistribution} provides a FOM very close to the maximum obtainable, even when the fiducial true cosmology is changed.}

\medskip

We also perform several tests to check the optimal distribution in a wide range of cases:
\begin{itemize}

\item[-]We check the stability of the shape of the optimal distribution by testing that small changes in the distribution still provide high \fom values.  We perform this test by finding the distributions that provide a \fom within 5\% of the maximum value. These distributions are calculated by sampling the \fom on a regular grid of the $n_{bin}$ parameters, using a step of $\Delta n_{bin}=10$. 
This check confirms that all the distributions corresponding to a high value of the \fom are similar to the one in  Table \ref{optdistribution}, i.e. have most of the clusters in the first, last and intermediate bin around $z\sim 0.42$. 

This is shown in Fig. \ref{distribution}, where we plot the optimal distributions for the $(w_0, w_a, \Omega_k=0)$ model, and the distributions with \fom within 5\% of the maximum \fom obtainable with the optimal distribution.

\begin{figure}[h!]
\begin{center}
\includegraphics[width=0.5\textwidth]{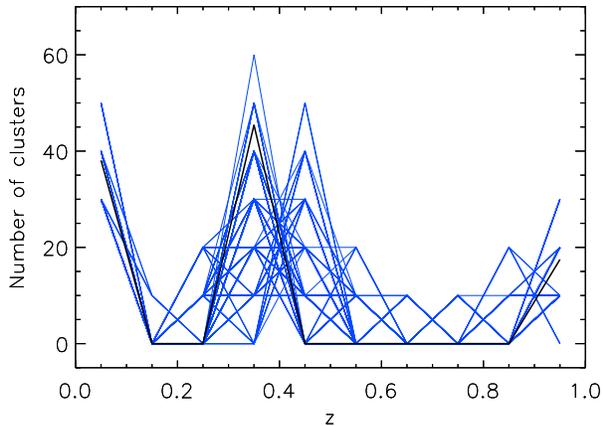}
\caption{The black thick line shows the optimal redshift distribution of galaxy clusters in 10 redshift bins in the range z=0-1, corresponding to the highest \fom=15.2 (from the Fisher Matrix analysis). The blue lines show the redshift distributions that have a \fom within 5\% of the maximum value.}
\label{distribution}
\end{center}
\end{figure}

\item[-]We check that the optimal distribution has roughly the same  behavior when different redshift ranges are considered, i.e., when the maximum redshift assumes a value between $z_{max}=0.9$ and $z_{max}=3$. This check has a purely theoretical interest for the higher redshifts, as it seems very unlikely that  present and near-future experiments  will detect many massive hot clusters at $z\gtrsim 1.5$. Nevertheless, we perform this analysis to better understand the shape of the optimal distributions. 
The optimal distributions appear to all have similar characteristics for different redshift ranges. Indeed,  all of them have around $\sim 35\%$ of the clusters in the first bin, $\sim 15\%$ in the last, and almost all the remaining clusters in the intermediate bin close to $z_{eq}$. A minor peak ($\sim 5\%$ of the clusters) also appears at $z\sim1.15$ when $z_{max}\gtrsim 1.8$.

Fig. \ref{figfomfisherw0w1fom} shows the constraints on the dark energy parameters, as well as the \fom, for a ($w_0-w_a,\Omega_k=0$) model, as a function of the maximum redshift $z_{max}$ of the redshift range $0-z_{max}$, for both the optimal and uniform distributions. 
It is interesting to note that the \fom obtained using a uniform distribution increases with the maximum redshift till $z_{max}=1.3$ and then starts decreasing again. The gain in \fom offered by pushing the observations of a uniform distribution to such a maximum redshift is, in any case, small as the \fom obtained is only 11.5.

\begin{figure}[h!]
\begin{center}
\includegraphics[width=0.45\textwidth,height=0.30\textheight]{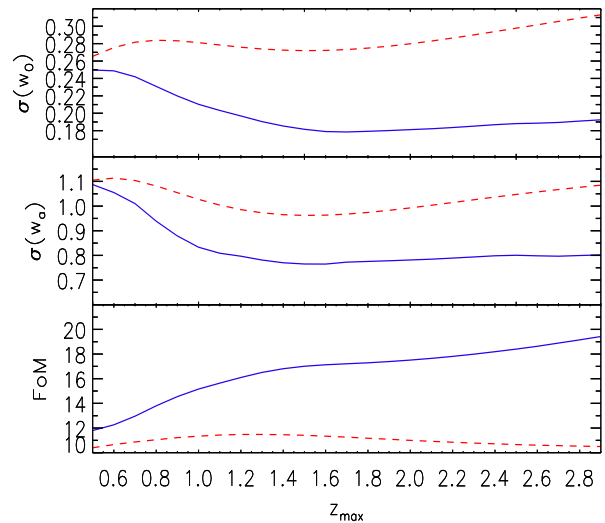}
\caption{\fom and the standard deviations of $w_0$ and $w_a$ as a function of the maximum redshift. We determine these constraints by combining simulated Planck CMB data and \fgas measurements for $100$ clusters in the redshift range $0-z_{max}$. We show both the results for an optimal (blue solid lines) and a uniform (red dashed lines) distribution of clusters.  These results are obtained with the Fisher Matrix analysis.}
\label{figfomfisherw0w1fom}
\end{center}
\end{figure}

\item[-]We also verify that a very similar  optimal distribution is obtained when instead of CMB data we consider priors of $\Omega_b h^2=0.0213\pm 0.002$ \cite{omeara} 
and $H_0=73.8 \pm 2.4$ \cite{riess11} in combination with the \fgas data. In this case we find that the optimal distribution is  almost the same as when using the full analysis with CMB+\fgas. This is due to the fact that now the clusters in the low redshift bin are used to constrain $\omega_c$ instead of $H_0$, as also this parameter has a strong effect on  the \fgas at low $z$.
For the $z=0-1$ range in the ($w_0-w_a$,$\Omega_k=0$) model, the maximum \fom achievable is only 3.9. This is due the absence of the high redshift measurement of the angular diameter distance provided by the CMB. In fact, while the effect of $w_0$ and $w_a$ on \fgas, i.e., on the ratio of the angular diameter distances $d_A^{\rm ref}/d_A^{\rm true}$, decreases with redshift some time after the epoch of equality, their integrated effect on the overall true angular diameter distance at high redshift is conserved. This is useful to constrain $w_a$, which has  a weak effect on \fgas (compared for example to $w_0$).
Thus, in absence of such a high redshift information, the \fom remains very small.

\item[-] We also study how much the systematics affect the optimal distribution and the \fom. The upper-left plot of Fig. \ref{optothermodels} shows the optimal distribution of Tab. \ref{optdistribution}, obtained assuming priors on systematics as in Tab. \ref{tab:systematics}, as well as the optimal distribution obtained assuming that systematic effects are perfectly known. In this last case, the last bin disappears, as the degeneracy between $H_0$ and the systematics is no longer present, while the intermediate peak of clusters shifts at the bin centered in $z=0.55$. The maximum \fom obtainable in this case is obviously much higher than in the case when systematic uncertainties are present, namely \fom$=51.5$. This indicates that the forecasted \fgas data is not enough to break degeneracies between systematics and cosmological parameters. Thus, priors heavily affect the constraints on dark energy.

\item[-]We check that the optimal distribution in Tab. \ref{optdistribution} for the ($w_0-w_a,\Omega_k=0$) model provides high \fom also when other theoretical frameworks are tested. We consider in total 4 models:
\begin{enumerate}
\item Linear redshift evolution of the dark energy equation of state in a flat universe ($w_0-w_a,\Omega_k=0$). This is the model we used to obtain the optimal distribution in Tab \ref{optdistribution}.
\item Linear redshift evolution of the dark energy equation of state with free curvature ($w_0-w_a,\Omega_k$).
\item Constant equation of state  in a flat universe ($w_0,w_a=0,\Omega_k=0$). 
\item Constant equation of state  with free curvature ($w_0, w_a=0,\Omega_k$).
\end{enumerate}
We find the optimal distribution for each of these models, as shown in Fig. \ref{optothermodels}. In the figure, we also show the optimal distribution one would obtain if all the systematics were perfectly known.
\begin{figure*}[t]
\begin{center}
\includegraphics[width=0.45\textwidth]{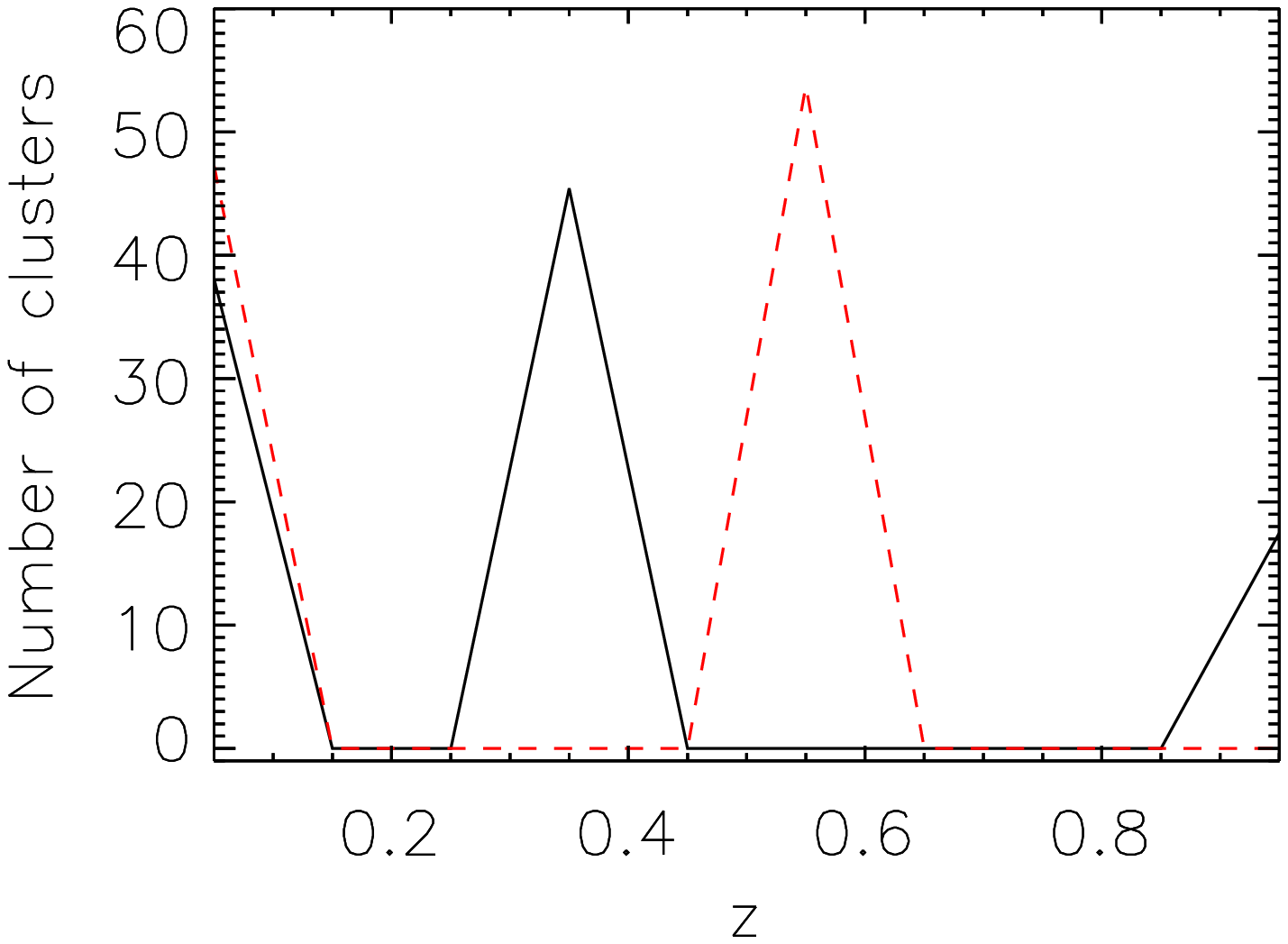}
\includegraphics[width=0.45\textwidth]{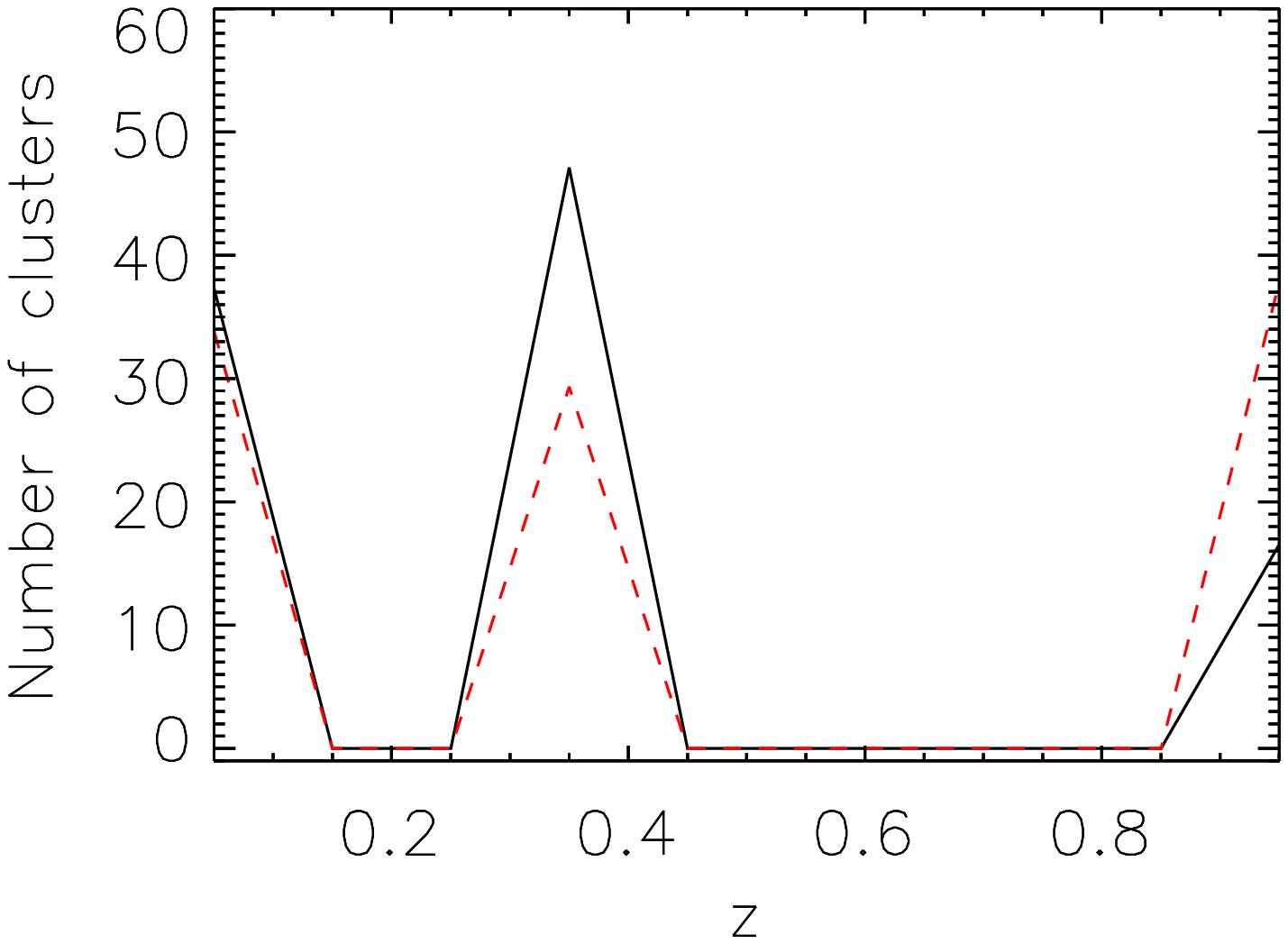}
\includegraphics[width=0.45\textwidth]{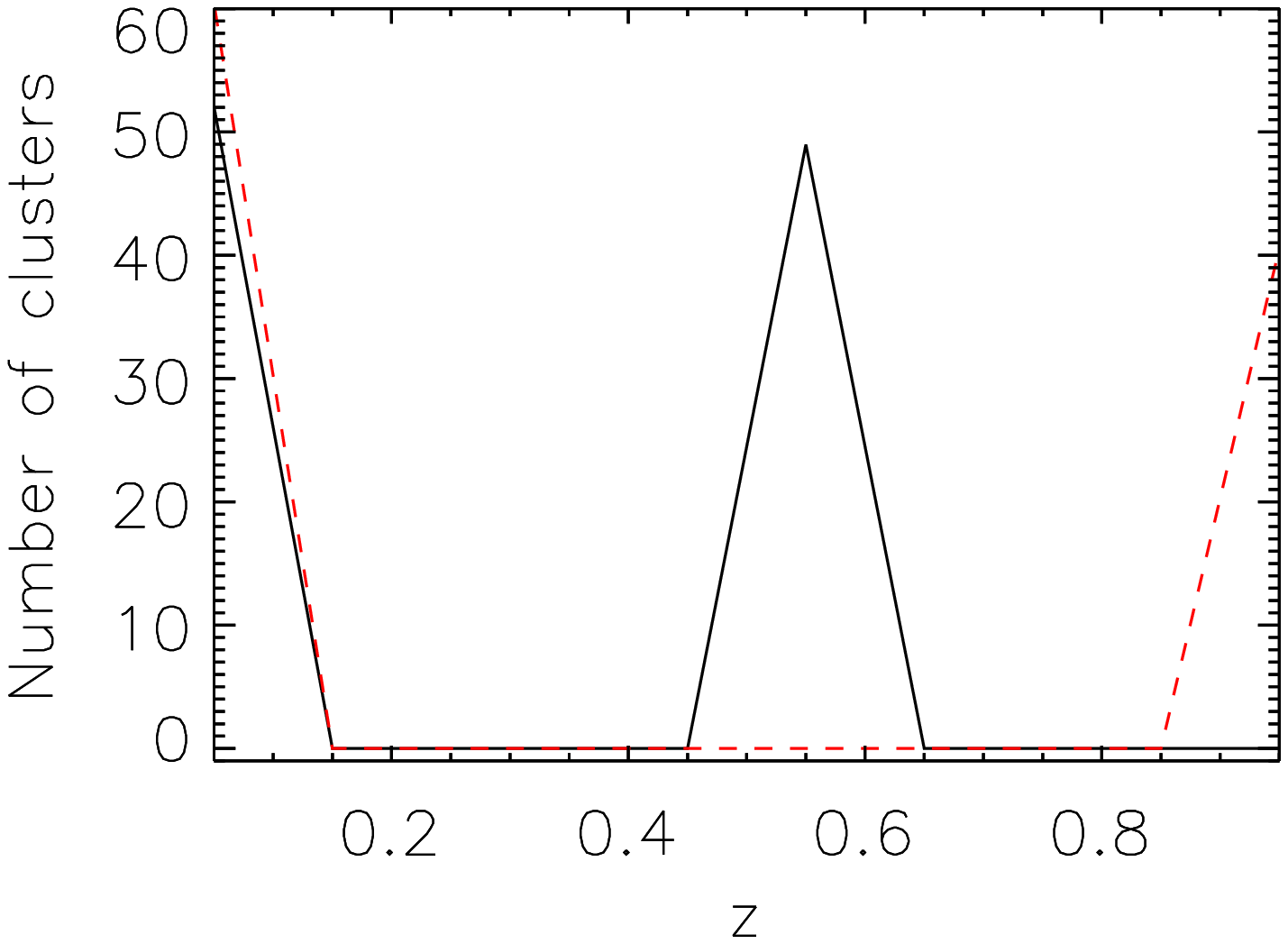}
\includegraphics[width=0.45\textwidth]{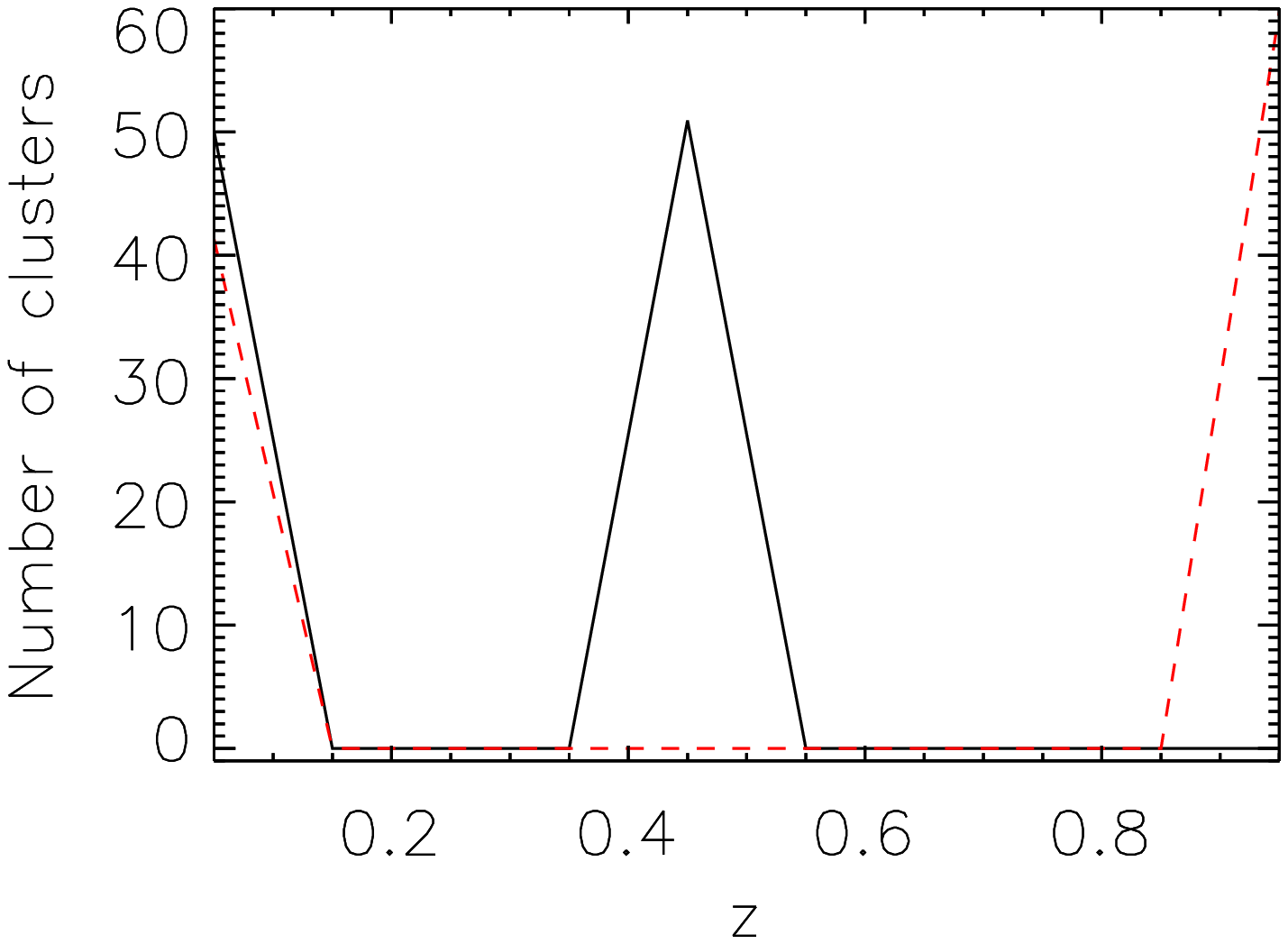}

\caption{Optimized distributions of clusters for the ($w_0$-$w_a$, $\Omega_k=0$) model (top-left), for ($w_0$-$w_a$, $\Omega_k$) (top-right), for ($w_0,w_a=0,\Omega_k=0$) (bottom-left) and for ($w_0w_a=0,\Omega_k$) (bottom-right). We show the results for the case using priors on systematics as in Tab.\ref{tab:systematics} (black solid lines) and assuming perfectly known systematics (red dashed lines).}
\label{optothermodels}
\end{center}
\end{figure*}

From the figure, it can be noticed that the distribution for the ($w_0-w_a,\Omega_k$) model is very similar to the one obtained in the flat case. The reason is that now the last redshift bin is used to determine curvature, rather than systematic effects. In this way, however, systematics are constrained only by the priors, lowering the possible \fom achievable, which is ${\rm FoM}=10.2$.

Considering now the case of a constant equation of state in a flat universe, $(w_0,w_a=0,\Omega_k=0)$, we find the optimal distribution by minimizing the standard deviation of $w_0$. In this case the optimal distribution presents only two peaks. Roughly half of the clusters are in the first bin, in order to determine the Hubble constant. The rest of the clusters are placed in an intermediate bin at $z=0.55$, where the effect of systematics is better disantagled from the effect of $H_0$. In fact, the uncertainty on the systematics grows with $z$, due to the redshift dependence of the stellar mass $a_s$ and of the depletion factor $\alpha_b$. However, $H_0$, measured at low redshift, is mostly degenerate with all those  systematic effects that are redshift independent, namely the combination of $K$, $\gamma$, $b_0$, and $s_0$. Thus, it is convenient to measure these systematics where the redshift-dependent effects are still subdominant, at redshift lower than 1. 
If  we now consider the case where  systematics are perfectly known, the peak at intermediate redshift disappears, while some clusters are placed in the last bin. This is due to the fact that measurements at this redshift can now improve the error bars obtained from the CMB on $\omega_c$ and $\omega_b$, and thus further decrease the degeneracy that these parameters have with the Hubble parameter at low redshifts in \fgas data.

Finally, we consider the case of a constant equation-of-state with free curvature, $(w_0,w_a=0,,\Omega_k)$. The optimal distribution  found is similar to the one for the flat case.

 We then compare the \fom and $\sigma(w_0)$ obtained with the optimal distributions in the different frameworks, with the constraints obtained using the optimal distribution of Tab. \ref{optdistribution} in the three other frameworks. The results are showed in Tab.\ref{fomfisher}. It is evident from the table that using the optimal distribution of Tab. \ref{optdistribution}  also in these other cases leads to \fom which are very close to the ones obtainable by using the proper optimal distribution for each case. For example, for the $(w_0,w_a=0,\Omega_k=0)$ model, the error obtained on $w_0$ using the proper optimal distribution is  $\sigma(w_0)=0.073$, while  the one obtained using the optimal distribution of Tab. \ref{optdistribution} yields to $\sigma(w_0)=0.079$; the variation is less than $\sim10$\%. Thus we conclude that the distribution in Tab. \ref{optdistribution} can be used in a wide range of models to obtain tight constraints on the dark energy equation-of-state.

\begin{table}[t]
\begin{center}
\begin{tabular}{  c| cccc}
\hline \hline
Model & \multicolumn{4}{c}{\fom}\\\hline
 &Uniform&\multicolumn{2}{c}{Optimal}&TAB. \ref{optdistribution} \\
& &Syst.& No syst.\\[5pt]
$w_0-w_a, \Omega_k=0$&11.3&15.2&51.5&.\\
$w_0-w_a,\Omega_k$&7.7&10.3&18.1&10.3\\[5pt]\hline
Model & \multicolumn{4}{c}{$\sigma(w_0)$}\\[5pt]\hline
 &Uniform&\multicolumn{2}{c}{Optimal}&TAB. \ref{optdistribution} \\
& &Syst.& No syst.\\[5pt]
$w_0,\Omega_k=0$&0.086&0.073&0.035&0.079\\ 
$w_0,\Omega_k$&0.11&0.098&0.048&0.10\\
\hline \hline
\end{tabular}
\caption{Constraints on \fom (or $\sigma(w)$ in the case of a constant dark energy equation of state) obtained by combining Planck CMB data and the \fgas data for 100 clusters with different distributions. We show the results using a uniform distribution, an optimal distribution calculated in each of the models considered and using the optimal distribution found for the ($w_0-w_a, \Omega_k=0$) case reported in Tab. \ref{optdistribution}. For the second distribution considered, we also report the results obtainable in the case of perfectly known systematic effects ('No syst.').}
\label{fomfisher}
\end{center}
 \end{table}
\end{itemize}

\section{MCMC analysis}
\label{section:mcmc}

\subsection{Method}
In order to check our previous results obtained using Fisher matrix forecasts, we derive the constraints on cosmological parameters using a modified version of
 the publicly available Markov Chain Monte Carlo package \texttt{cosmomc} \cite{Lewis:2002ah}.
We use a convergence
diagnostics based on the Gelman and Rubin statistic performed on $4$ chains, and the Parameterized Post-Friedmann(PPF) prescription for the dark energy perturbations \cite{husawicki} implemented by \cite{fang} in the \texttt{camb} code. Moreover, we use the code of \cite{rapetticode} to implement the \fgas data in the \texttt{cosmomc} code.

We simulate datasets for the Planck experiment 
following the commonly used approach, described, 
for example, in \cite{perotto}. We first generate the fiducial theoretical angular power
spectra $C_{\ell}^{TT}$, $C_{\ell}^{TE}$, $C_{\ell}^{EE}$ with the \texttt{camb}  code for a cosmological model compatible with the WMAP5 maximum likelihood parameters reported in Table \ref{tab:fiducial}. We then add the experimental noise $N_\ell$   described in Equation \ref{noisecmb}.

For \fgas we simulate a dataset of 100 clusters over the redshift range $z=0-1$, distributed with either the optimal configuration reported in Table \ref{optdistribution} or with a uniform distribution. We assume that the true and the reference cosmology are the same as the fiducial model of Table \ref{tab:fiducial}. A 10\% error is assigned to the measurements.

We sample a set of cosmological parameters  similar to the one used in the Fisher Matrix analysis, adopting flat priors: the physical baryon and cold dark matter densities, $\omega_b=\Omega_bh^2$ and $\omega_c=\Omega_ch^2$, the ratio of the sound horizon to the angular diameter
 distance at decoupling, $\theta_S$ (used instead of $H_0$), and the curvature $\Omega_k$. For the CMB data, we additionally consider the scalar spectral index $n_S$, the overall normalization of the spectrum  $A$ at $k=0.002$ Mpc$^{-1}$, and the optical depth to reionization, $\tau$. Furthermore, for \fgas we consider the seven systematic parameters described in Section \ref{systematics2}, with the priors in Table \ref{tab:systematics}.

We test the four models considered in the Fisher Matrix analysis, i.e., a linear evolution of the dark energy equation of state $w_0-w_a$ in a flat and non-flat universe, and a constant value of the dark energy equation-of-state $w_0$ in a flat and non-flat universe.
 
\subsection{Results}

Table \ref{fomMCMC} shows our constraints on the most interesting varied and derived parameters for all the models and  combinations of data tested, obtained using the \texttt{cosmomc} code. 
Concerning the $(w_0-w_a, \Omega_k=0)$ model, the \fom found with a combination of the synthetic Planck CMB data plus an optimal distribution of \fgas data is smaller than the Fisher Matrix value in Table \ref{fomfisher} ($14.4$, instead of $15.1$) due to better accounting of uniform priors and a more reliable treatment of the non-Gaussian probability distributions of the parameters. 
Nevertheless, the combination of Planck CMB+\fgas data with an optimal distribution still improves the \fom by $\sim26\%$ 
compared to a uniform distribution over the same redshift range, which provides a \fom of 11.4. 
\begin{table*}[h!t]
\begin{center}
\begin{ruledtabular}
\begin{tabular}{  lccccccc}
\noalign{\vskip 5pt}
 Data                           &$\sigma(w_0)$                &$\sigma(w_a)$     &$\sigma(\Omega_{de})$           &$\sigma(\Omega_m)$                &$\sigma(H_0)$ &$\sigma(\Omega_K)$& FOM \\ \noalign{\vskip 5pt}\hline

\noalign{\vskip 5pt}
\multicolumn{8}{c}{ Model: $w_0-w_a, \Omega_k=0$} \\\noalign{\vskip 5pt}
\hline 

\noalign{\vskip 5pt}
WMAP5+Allen \fgas   &                                                               0.28&                      0.97&                      0.035&                      0.035&                      4.3&.&  8.3
\\[2 pt]
Planck+\fgas (opt.)   &                                                             0.20&                      0.75&                      0.028&                      0.028&                      3.5&.& 14.5
\\[2 pt]
Planck+\fgas (uni.)   &                                                             0.25&                      0.88&                      0.028&                      0.028&                      3.7&.& 11.4
\\[2 pt]
Planck+ Allen \fgas   &                                                             0.28&                      1.0&                      0.029&                      0.029&                      4.1&.& 8.8
\\[2 pt]
Planck+   &                                                                        0.17&                      0.54&                      0.024&                      0.024&                      3.0&.& 24.2
\\[2 pt]
+\fgas[Allen + (opt.)]\\[2pt]
\hline
\noalign{\vskip 5pt}
\multicolumn{8}{c}{ Model: $w_0-w_a, \Omega_k$} \\\noalign{\vskip 5pt}
\hline 

\noalign{\vskip 5pt}

Planck+\fgas (opt.)   &                                                      0.20&                      0.94&                      0.031&                      0.040&                      4.6&  0.016&7.5 
\\[2 pt]
Planck+\fgas (uni.)   &                                                      0.26&                      1.0&                      0.032&                      0.041&                      4.7&  0.014&6.1 
\\[2 pt]

\hline
\noalign{\vskip 5pt}
\multicolumn{8}{c}{ Model: $w_0,w_a=0, \Omega_k=0$} \\\noalign{\vskip 5pt}
\hline 
\noalign{\vskip 5pt}
Planck+\fgas (opt.)   &                                                           0.091&                      .&                      0.026&                      0.026&                      3.3&.&$.$
\\[2 pt]
Planck+\fgas (uni.)   &                                                           0.098&                      .&                      0.028&                      0.028&                      3.5&.&$.$

\\[2 pt]
\hline
\noalign{\vskip 5pt}
\multicolumn{8}{c}{ Model: $w_0,w_a=0, \Omega_k$} \\\noalign{\vskip 5pt}
\hline 

\noalign{\vskip 5pt}

Planck+\fgas (opt.)   &                                                          0.14&                      .&                      0.030&                      0.039&                       4.6&0.013&$.$
\\[2 pt]
\end{tabular}
\end{ruledtabular}
\caption{Figures of merit and 68\% c.l. errors on cosmological parameters for different models, obtained using the \texttt{cosmomc} code. The dark energy and dark matter density parameters $\Omega_{de}$ and $\Omega_{m}$ are indirectly derived.
We show the results for different combinations of data. `WMAP5' indicates CMB data from the 5-year release of the WMAP satellite data \cite{wmap5}, while `Planck' indicates simulated  CMB data from the Planck satellite \cite{planck}.
`\fgas (opt).' indicates \fgas simualated data for $100$ clusters with an optimal distribution between redshift $0$ and $1$, while '\fgas (uni.)' is for a uniform distribution. Finally, `Allen' indicates \fgas real data for 42 clusters publicated in Allen et al. \cite{allen}.}\label{fomMCMC}
\end{center}
 \end{table*}

Moreover, the results obtained with the combination of synthetic Planck CMB+optimal \fgas represent a major improvement over current data. For instance, we calculated the \fom from WMAP5 CMB combined with the \fgas data for $42$ clusters measured by Allen et al.\cite{allen}. These datasets yield a \fom of 8.3; therefore, future data will improve this result by almost a factor $\sim 2$, as clear from the second line of Tab. \ref{fomMCMC}. An even more impressive improvement is obtained if in addition to the synthetic Planck CMB data and the \fgas data for 100 clusters with an optimal distribution, we add the known 42 \fgas measurements of Allen et al. In this case the \fom will increase of almost a factor $3$.

The results in Table \ref{fomMCMC} also demonstrate that the improvement in the determination of the dark energy equation-of-state will be mainly driven by more accurate \fgas data rather than better CMB data. In fact, combining Planck CMB data with Allen et al.'s \fgas measurements only provides a \fom slightly better than that the one obtained with WMAP5 data --- nominally \fom=8.8.

\begin{figure}[th]
\begin{center}
\includegraphics[width=0.45\textwidth]{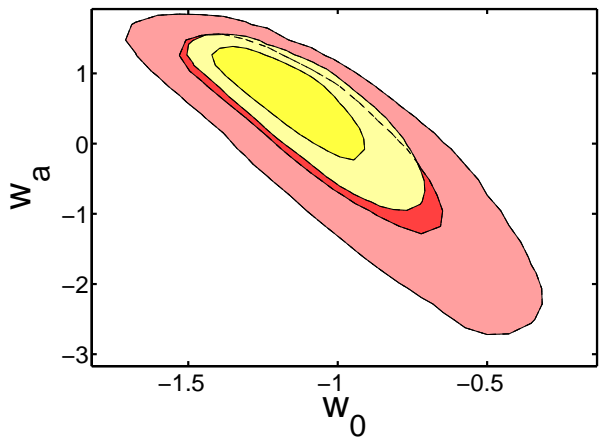}
\caption{68\% and 95\% likelihood contour plots on the $w_0$ - $w_a$ 
 plane for WMAP5 plus \fgas data for the 42 clusters reported in Allen et al. \cite{allen} (red) and synthetic Planck plus \fgas data for the 42 clusters reported in Allen et al.\cite{allen}, plus \fgas data for 100 clusters between redshift 0 and 1 with an the optimal distribution (yellow). Here the model considered is $(w_0-w_a,\Omega_k=0)$ in a flat universe.}
\label{2Dw0wawmap}
\end{center}
\end{figure}

\begin{figure}[th]

\begin{center}
\includegraphics[width=0.49\textwidth]{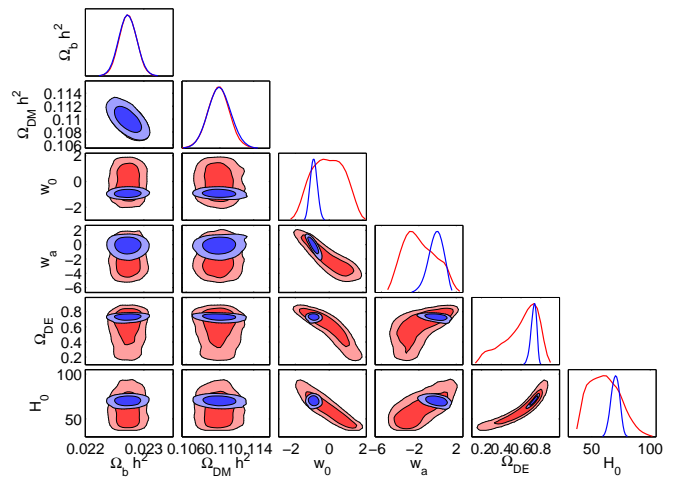}
\caption{Constraints for the $(w_0-w_a,\Omega_k=0)$ model from synthetic Planck CMB data (red) and CMB+ \fgas data (blue) for 100 clusters with an optimal distribution between $z=0-1$. We show the marginalized one-dimensional distributions and two-dimensional 68\% and 95\% c.l. contour plots.}
\label{2Dtri_w0w1}
\end{center}
\end{figure}

Fig. \ref{2Dw0wawmap} shows the 68\% and 95\% likelihood contour plots  in the $w_0$ - $w_a$ plane, obtained using current data and synthetic Planck CMB plus \fgas data for the 42 clusters reported in Allen et al.\cite{allen} plus \fgas  forecasted data with an optimal cluster distribution. Fig. \ref{2Dtri_w0w1} compares the constraints achievable by Planck CMB only and  Planck CMB+\fgas forecasted data for 100 clusters with an optimal distribution.
In this plot, the degeneracy-breaking power of the \fgas data on the parameters that  determine the geometry of the universe, such  as the dark energy equation of state parameters $w_0$, $w_a$ and the Hubble constant, is striking.  By themselves, CMB data can only place weak constraints on these parameters when they are considered simultaneously.
As expected, adding \fgas data to CMB data do not improve the constraints on other parameters, namely the baryon and dark matter densities 
$\omega_b$ and $\omega_c$, the spectral index $n_s$, and the optical depth $\tau$.

\begin{figure}[th]
\begin{center}
\includegraphics[width=0.49\textwidth]{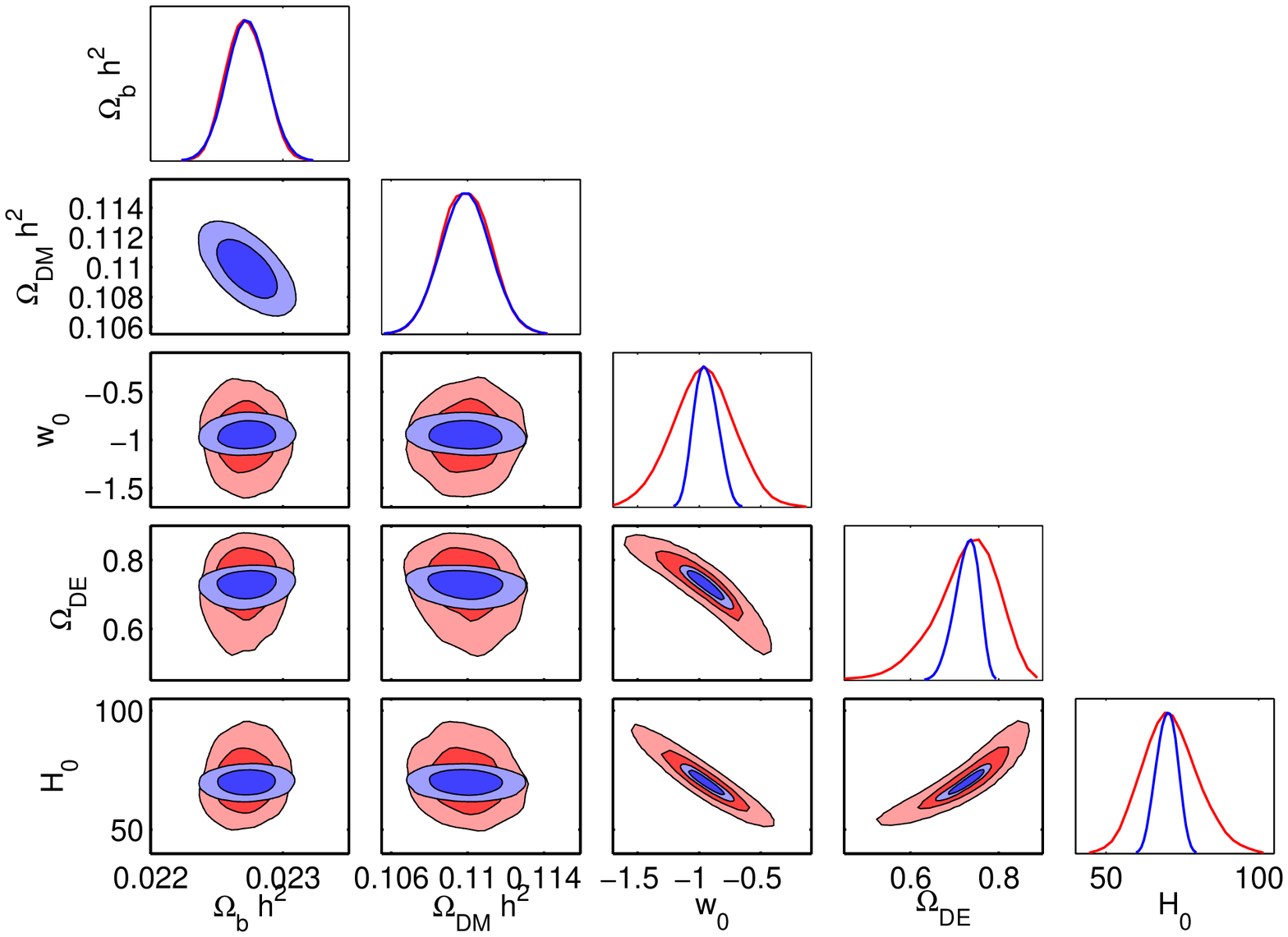}
\caption{Constraints for the $(w_0, w_a=0,\Omega_k=0)$ model from synthetic Planck CMB data (red) and CMB+ \fgas data (blue) for 100 clusters with an optimal distribution between $z=0-1$. We show the marginalized one-dimensional distributions and two-dimensional 68\% and 95\% c.l. contour plots.}
\label{2Dtri_w}
\end{center}
\end{figure}

Very similar considerations can be made for a constant equation of state $w=w_0$ in a flat universe ($\Omega_k=0$), 
as show in Figure \ref{2Dtri_w}. 
In this case, the constraint on $w_0$ is $\sigma(w_0)=0.091$, which represents a $35\%$ improvement compared to the results reported in Allen et al. using WMAP3 CMB data and 42 clusters, namely $\sigma(w_0)=0.14$. 

\begin{figure}[th]
\begin{center}
\includegraphics[width=0.49\textwidth]{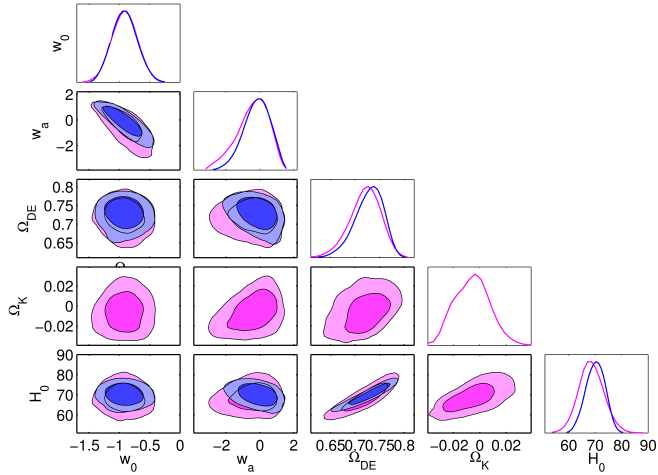}
\caption{Constraints for the $(w_0-w_a,\Omega_k)$ model from synthetic Planck CMB data + \fgas data (magenta) for 100 clusters with an optimal distribution between $z=0-1$, compared to constraints for a flat universe (blue).
We show the marginalized one-dimensional distributions and two-dimensional 68\% and 95\% c.l. contour plots.}
\label{2Dtri_w0w1omk}
\end{center}
\end{figure}

Finally we repeated the analysis for also the case of a universe with free curvature. Fig. \ref{2Dtri_w0w1omk} shows the constraints obtained for the $(w_0-w_a,\Omega_k)$ model. 
 The main effect of adding free curvature is to introduce more degeneracy with the parameters of interest. This leads to a widening of the errors and a distortion of the probability distributions. This also explains why the \fom obtained with the MCMC analysis in $(w_0-w_a,\Omega_k)$ case, \fom=$7.5$, is 30\% smaller than the one estimated with the Fisher Matrix, \fom=10.3, that assumes Gaussian likelihoods for the considered parameters.

\section{Conclusions}
\label{sec:conclusions}

We consider constraints on the dark energy equation-of-state obtained when combining the future CMB data expected from the Planck satellite and simulated \fgas measurements for 100 clusters observed by a possible future X-ray campaign with X-ray satellites.  We found the optimal cluster distribution over the redshift range $z=0-1$ to provide the tightest constraints on the dark energy equation-of-state.  The optimal distribution improves the \fom by almost $\sim 30\%$ compared to a uniform distribution. We found that the optimal distribution, found for a linearly evolving dark energy equation of state in a flat universe, provides very high figures of merit also in other theoretical frameworks, namely free curvature or a constant equation-of-state.  A positive characteristic of the optimal distribution is that clusters are placed at low and intermediate redshifts, while only a few should be observed at high redshift, where measurements require longer observational times.
We performed our analysis by both using the Fisher Matrix and the MCMC methods to forecast constraints.
We found that future data will increase the \fom by a factor $\sim 2$ relative to the current constraints available with WMAP5 and the \fgas data from \cite{allen}.  Our study provides a useful guide for planning XMM-Newton and Chandra follow-up observations of Planck and SPT clusters. 
\acknowledgments
S.G. would like to thank D. Rapetti, B. Wandelt and S. Allen for useful discussions.  A
portion of the research described in this paper was carried out at the Jet
Propulsion Laboratory, California Institute of Technology, under a contract with
the National Aeronautics and Space Administration. This work was supported by the PRIN-INAF grant 'Astronomy probes
fundamental physics.

 \newcommand\MNRAS[3]{Mon. Not. R. Astron. Soc.{\bf ~#1}, #2~(#3)}

\end{document}